\documentclass{ws-rv9x6}
\usepackage{graphicx}
\makeindex
\begin{document}

\chapter{Pairing at High Spin}\label{ra_ch1}

\author{S. Frauendorf\footnote{sfrauend@nd.edu}}

\address{Department of Physics, University Notre Dame, IN 37556, USA \\}

\begin{abstract}
Pair correlations are described  in the framework of the HFB approximation applied to a 
uniformly rotating system (Cranking model). The reduction of the moments of inertia, the 
classification of rotational bands as multi quasiparticle configurations, and the signatures of the 
rotation induced transition to the unpaired state are discussed.     
\end{abstract}
\body

\section{Superfluidity 
vs. nuclear pairing}\label{s:super}
One of the  fundamental characteristics of a the superfluid  is  its irrotational flow pattern in a rotating
container that is deformed with respect to the axis of rotation. 
In contrast, a viscous fluid develops rigid rotational
flow with the velocity field $\vec v=\vec \omega \times \vec r$. 
Irrotational flow is the direct consequence of the fact that the superfluid state 
is described by the coherent wave function $\Psi(\vec x)= \left|\Psi(\vec x)\right|\exp[i\theta(\vec x)]$ of the order parameter.
The condensate density $\rho=\left|\Psi(\vec x)\right|^2$ is constant except in a very thin layer near the container wall. Hence the mass current inside the container is 
$\vec j=  \mathbf{Im}\left[\Psi(\vec x)^*\vec p \Psi(\vec x)\right]=\rho\hbar \nabla \theta(\vec x)$,
meaning that $\vec v=\nabla \theta(\vec x)$  is irrotational. The moment of inertia of irrotational flow ${\cal J}_{irrot}$ is substantially smaller than the rigid body moment of inertia ${\cal J}_{rig}$. 
As a consequence, the superfluid phase becomes energetically disfavored to the normal phase at the critical angular velocity 
$\omega_c$, where
\begin{equation}\label{e:omc}
E_P-\frac{\omega_c^2}{2}{\cal J}_P=E_N-\frac{\omega_c^2}{2}{\cal J}_N.
\end{equation}
Here $E_P$ and $E_N$ are  the energies of the liquid in the superfluid (paired) and normal state, respectively, and ${\cal J}_P={\cal J}_{irrot}$ and
${\cal J}_N={\cal J}_{rig}$ are the respective moments of inertia . The phase transition is of first order. It sets in
at the angular momentum $J=J_{cP}=\omega_c{\cal J}_P$, where part of the liquid becomes normal. For  $J_{cP}<J<J_{cN}$ it is in the mixed state, where
 the normal phase  coexists
 in the form of quantized vortices. At $J=J_{cN}=\omega_c{\cal J}_N$ the liquid is completely normal (see inset of Fig. \ref{f:MoInop}).

\begin{figure}[t]
\centerline{\includegraphics[scale=0.5]{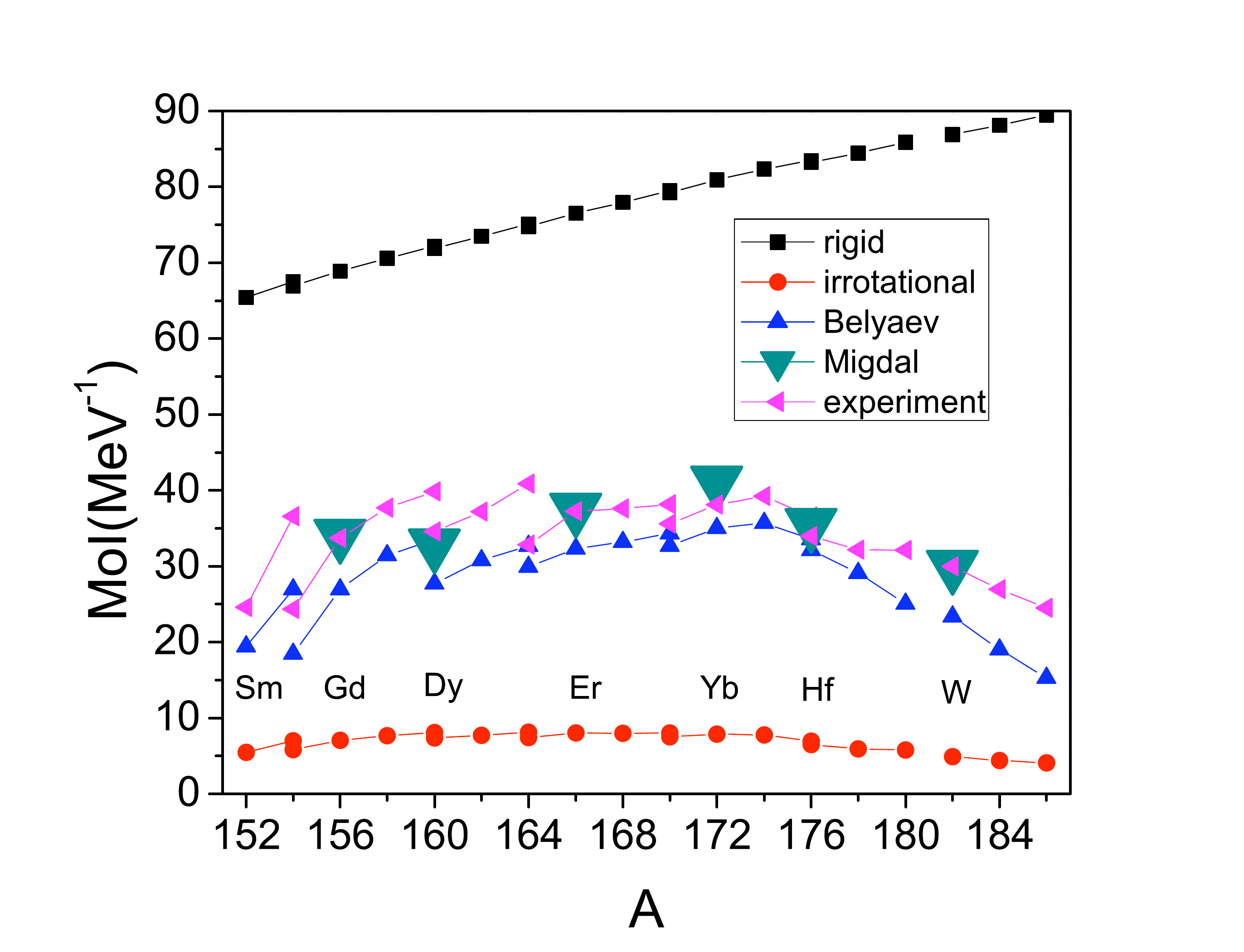}}
\caption{Moments of inertia of rare earth nuclei. "Belyaev" shows the calculation by Nilsson and Prior 
\cite{NilssonPrior} using Belyaevs's \cite{Belyaev} expression "Migdal" shows the calculation 
by Hamamoto \cite{Hamamoto} using Migdal's expression \cite{Migdal}. } \label{f:MoI}
\end{figure}

The experimental values of the moments of inertia in Fig. \ref{f:MoI} are only about one half of the rigid body value.  Bohr, Mottelson and 
Pines \cite{BohrMottPines} suggested that pair correlation like the ones  causing superconductivity in metals are responsible
 for the reduction. However the fact that they
are about six times larger than the irrotational value indicates a fundamental difference between nuclei and  a macroscopic ideal liquid.

 Microscopically, the rotating superfluid is described by the Hamiltonian in the rotating frame $H'$, also called the Routhian, 
\begin{equation}\label{e:routhian}
H'=H_N+V_P-\vec \omega \cdot \vec J, ~~~\vec J=\vec L+\vec S,
\end{equation}
where $\vec L$ is the total orbital angular momentum, $\vec S$ the total spin of the constituents, $H_N$ the Hamiltonian
of the normal state and $V_P$ the attractive interaction that causes superfluidity. The BCS approximation to this many body
problem leads to the coherence length $\xi=\hbar p_F/(\pi m \Delta)$, which measures the size of a Cooper pair. The 
macroscopic condensate wave function $\Psi(\vec x)$ and the ensuing irrotational flow emerge on a length scale that is large compared to $\xi$. In the surface layer of width $\xi$ the flow deviates from being irrotational. In the case of nuclei, with the gap parameter $\Delta =1$ MeV,  the Fermi momentum $\hbar/p_F=0.75$ fm, and the Fermi energy $e_F=37$ MeV, one finds $\xi=2/\pi( \hbar/p_F)(e_F/\Delta)=18$ fm, which is substantially larger than the size of the nucleus. Although it is quite common
to say that nuclei are superfluid at  low spin, this is not really appropriate. 
Nuclei are made of the surface layer of a superfluid, so to speak. Another difference is that  the Cooper pairs  are composed of nucleons on quantized orbitals that
carry distinctly different angular momentum (high-j and low-j orbitals).    
The consequences of pairing for nuclear rotation  can only be described by means of a microscopic approach. The BCS theory is 
often sufficient. However, for example, for understanding the rotation-induced transition to the unpaired state one has to go beyond the BCS mean field approximation.  
    
\section{Cranked Hartree-Fock-Bogoliubov approximation}\label{s:CHFB}

The microscopic description of pairing at high spin applies the Hartree-Fock-Bogoliubov (HFB) approximation to
the Routhian (\ref{e:routhian}). It is called "Cranked HFB" (CHFB), because the system is put into a uniformly rotating
(cranked) frame of reference. The HFB is discussed in detail by Dobaczewski and Nazarewicz in this book. 
We consider the most common  case that the nuclear potential is reflection symmetric and the rotational
axis $\vec \omega$ coincides with one of the principal axes (x- )  of the potential.
The quasiparticle (qp) creation operators   are generated by the Bogoliubov transformation, 
\begin{equation}\label{e:Btransf}
\beta^+_{\alpha \mu}=\sum_{i}U_{\alpha i}^{\alpha \mu} c^+_{\alpha i}+V_{ -\alpha i}^{ \alpha \mu} c_{-\alpha i}.
\end{equation}
The amplitudes $U_{\alpha i}^{\alpha \mu},~V_{ -\alpha i}^{ \alpha \mu}$ and the qp  energies $e'_{\alpha \mu}$ are, respectively,
 the eigenvectors and eigenvalues of the qp  Routhian 
\begin{equation}\label{e:qpHam}
{\cal H}_{\alpha i,\alpha k}^\omega=\left[ \begin{array}{cc} \left(h_{sp}-\omega j_x-\lambda \right)_{\alpha i,\alpha k} & \Delta_{\alpha i,-\alpha k } \\
 \Delta_{-\alpha i,\alpha k}  &- \left(h_{sp}-\omega j_x-\lambda \right)_{-\alpha i,-\alpha k} \end{array} \right],
\end{equation}
\begin{equation}\label{e:qpEvEq}
\sum_{k}{\cal H}_{\alpha i,\alpha k}^\omega\left[ \begin{array}{c} U_{\alpha k}^{\alpha \mu} \\ V_{ -\alpha k}^{ \alpha \mu} \end{array} \right]
=e'_{\alpha \mu}\left[ \begin{array}{c} U_{\alpha i}^{\alpha \mu} \\ V_{ -\alpha i}^{ \alpha \mu} \end{array} \right].
\end{equation}
The single particle hamiltonian $h_{sp}$ derives from the effective interaction in the particle-hole channel via additional the selfconsistency conditions  (see section \ref{s:meanfield}). Here we consider it as given.   
As we assume it is reflection symmetric,  the parity $\pi$ is good for
the single particles states and for the qps.  The single particle Routhian $h_{sp}-\omega j_x$ is invariant with respect to a rotation 
${\cal R}_x$ by $\pi$ about the x-axis, which implies the signature quantum number $\alpha$ for the single particle states and the qps,
\begin{equation}
{\cal R}_xc^+_{\alpha i}{\cal R}_x^{-1}=e^{-i\alpha \pi}c^+_{\alpha i},~~~{\cal R}_x\beta^+_{\alpha i}{\cal R}_x^{-1}=e^{-i\alpha \pi}\beta^+_{\alpha i},~~~\alpha=\pm \frac{1}{2}.
\end{equation}
The states with opposite signature are related by time reversal, i. e. the Cooper pairs are composed of nucleons of opposite signature. 
For each solution $\alpha \mu$ there is a conjugate solution $-\alpha \mu$ with $E_{-\alpha \mu}=-E_{\alpha \mu}$ 
and $U_{\alpha i}^{-\alpha \mu}= V_{ -\alpha i}^{ \alpha \mu}$, $V_{-\alpha i}^{-\alpha \mu}= U_{ \alpha i}^{ \alpha \mu}$.

The pair field $ \Delta_{\alpha i,-\alpha k}$ is obtained by the selfconsistency condition 
\begin{equation}
 \Delta_{\alpha i,-\alpha k} =\frac{1}{2}\sum_{i' k' \alpha' \mu}g_{\alpha i,-\alpha k;\alpha' k'-\alpha' i'}U_{\alpha', k'}^{\alpha' \mu} 
 V_{ -\alpha' i'}^{ \alpha' \mu} n_{\alpha' \mu}\label{e:ScDel}
\end{equation}from the pairing interaction 
\begin{equation}
V_P=-\frac{1}{4}\sum_{ii'kk' \alpha \alpha'}g_{\alpha i,-\alpha k;\alpha' k',-\alpha' i'}c^+_{\alpha i}c^+_{ -\alpha k}c_{ -\alpha' k'}c_{\alpha' i'}.
 \end{equation}
The qp  occupation numbers $n_{\alpha \mu}$  determine a certain qp configuration (see section \ref{s:elem}).

The  CHFB treats nuclear rotation in a semiclassical way, which is an accurate approximation 
for rotational bands composed of a long sequence of states of discrete angular momentum eigenvalues $I$. The contact to experiment is made either by 
the constraint that for a given qp  configuration the angular momentum expectation value $\langle j_x\rangle=J=\sqrt{I(I+1)}\hbar\approx( I+1/2)\hbar$ or by 
directly referring to the experimental angular frequency (see section \ref{s:elem}).   The classical canonical relations 
\begin{equation}\label{e:can}
E'=E-\omega J,~~~ \frac{dE(J)}{dJ}=\omega,~~~J=-\frac{dE'(\omega)}{d\omega} 
\end{equation}
between the energy $E$ in the laboratory frame, the energy $E'$ in the rotating frame (Routhian), and the angular momentum $J$ hold
for CHFB and turn out important for the interpretation.  

\section{Perturbative solutions - Moments of inertia}\label{s:MoI}

For $\omega=0$,  Eqs. (\ref{e:qpHam}-\ref{e:ScDel}) describe the pair correlations in nuclei carrying no angular momentum.
Belyaev \cite{Belyaev} first derived an expression for the moments of inertia ${\cal J}$ by taking into account the "cranking term" $-\omega j_x$
in linear perturbation theory. In obtaining his expression,   he used the  "monopole" pairing interaction  
\mbox{$g_{\alpha i,-\alpha k;\alpha' k',-\alpha' i'}=G\delta_{ik}\delta_{i'k'}$}, which gives a state-independent pair field $\Delta$ (gap parameter).
 He estimated ${\cal J}/{\cal J}_{rig}\sim1/2$. A more realistic calculation 
based on Belyaev's expression was carried out by 
Nilsson and Prior \cite{NilssonPrior}, which is shown in Fig. \ref{f:MoI}. They used  Nilsson's modified oscillator potential for $h_{sp}$ and carefully
adjusted the deformation parameter to the measured electric quadrupole moments  and the pairing gaps $\Delta$ to the experimental 
even-odd mass differences. The calculated values of ${\cal J}$ follow closely the experimental ones, but are systematically somewhat too small.  
Migdal \cite{Migdal} derived a microscopic expression for the moment of inertia using a more general expression for the 
pairing interaction that obeyed local Galilean invariance. He obtained a modification of the  pair field via the selfconsistency condition (\ref{e:ScDel}),
which is linear in $\omega$, and a correction term to Belyaev's expression. This so called Migdal term 
ensures that ${\cal J}\rightarrow{\cal J}_{irrot}$ for $\Delta\rightarrow\infty$.  A quantitative 
 evaluation of the Migdal term was given by Hamamoto \cite{Hamamoto}, who  calculated the moments of inertia in a similar way as in
Ref. \cite{NilssonPrior}. In order to appproximately restore the Galilean invariance, she complemented the  
 monopole pairing interaction by the  quadrupole pairing,  
\mbox{$g_{\alpha i,-\alpha k;\alpha' k',-\alpha' i'}=G_0\delta_{ik}\delta_{i'k'}+G_2\sum_\nu(-)^\nu q^\nu_{\alpha i -\alpha k}q^{-\nu}_{-\alpha' i' \alpha'k}$}.
Her results in Fig. \ref{f:MoI} agree well with the experimental moments of inertia. The Migdal term amounts to about 15\% of the total.   

The next order of the perturbation series gives the total angular momentum and Routhian as
\begin{eqnarray}\label{e:ref}
J(\omega)=\langle j_x\rangle=\omega{\cal J}_0+\omega^3{\cal J}_1,\\
E'(\omega)=\langle H'\rangle=E(0)-\frac{\omega^2}{2}{\cal J}_0-\frac{\omega^4}{4}{\cal J}_1,
\end{eqnarray}
where ${\cal J}_0$ is the moment of inertia just discussed. 
Marshalek \cite{Marshalek} worked out the expressions for ${\cal J}_1$ and calculated them along the lines of Ref. \cite{NilssonPrior}. He found that 
it arises to about equal parts from the reaction of the qps to the Coriolis force (see section \ref{s:elem}) and the attenuation of $\Delta$,
called Coriolis Antipairing (CAP). His ${\cal J}_1$ values turned out to be systematically too large compared to experiment. Frauendorf \cite{Frauendorf} noticed that
the CHFB overestimates the CAP.  Including projection onto good particle number resulted in a substantially reduced CAP, which reconciled the
${\cal J}_1$ values with experiment. Note, the  CAP does not appear in a macroscopic super fluid, because the moment of inertia takes the irrotational 
value irrespective of the value of $\Delta$.

\section{Rotating quasiparticles - the Cranked Shell Model}\label{s:elem}\label{s:CSM}

Banerjee, Mang, and Ring \cite{Banerjee}  found the  first non-perturbative solution of the CHFB equations  for a monopole pairing
interaction. As will be discussed in section \ref{s:trans}, the CAP turned out to be modest for a given qp  configuration. 
This allowed Bengtsson and Frauendorf \cite{BFqp} (BF) to interpret the lowest rotational bands as qp  configurations in a rotating
deformed potential with a fixed pair field $\Delta$.  Their approach, called Cranked Shell Model (CSM),  aims only 
at relative energies and angular momenta, which turned out to
be rather well  accounted for by configurations of weakly interacting qps. Since the reaction of the qps to the inertial forces is determined 
by the angular frequency $\omega$, the decisive step was  plotting the Routhians $E'$  and angular momenta $J$ of the qp configurations  
as functions of $\omega$ relative to the reference values $E'_g$ and $J_g$ of the qp vacuum.
They introduced the experimental rotational frequency $\omega(I)=\left(E(I+1)-E(I-1)\right)/2$, which is one half of the frequency of the
quadrupole radiation, as expected from semiclassics. The experimental functions $J(\omega)$ and $E'(\omega)$
were obtained according to Eqs. (\ref{e:can}) by  interpolating between the discrete points.  The vacuum values $E'_g$ and $J_g$ were
 parametrized by the "Harris reference" (\ref{e:ref}), where ${\cal J}_0$ and ${\cal J}_1$ were adjusted to the experimental energies of the 
 ground state (g-) rotational band of the considered even-even nucleus.

\begin{figure}[t]
\begin{minipage}[b]{0.58\linewidth}
\centering
\vspace*{-0.5cm}
\hspace*{-1.6cm}
\includegraphics[scale=0.39]{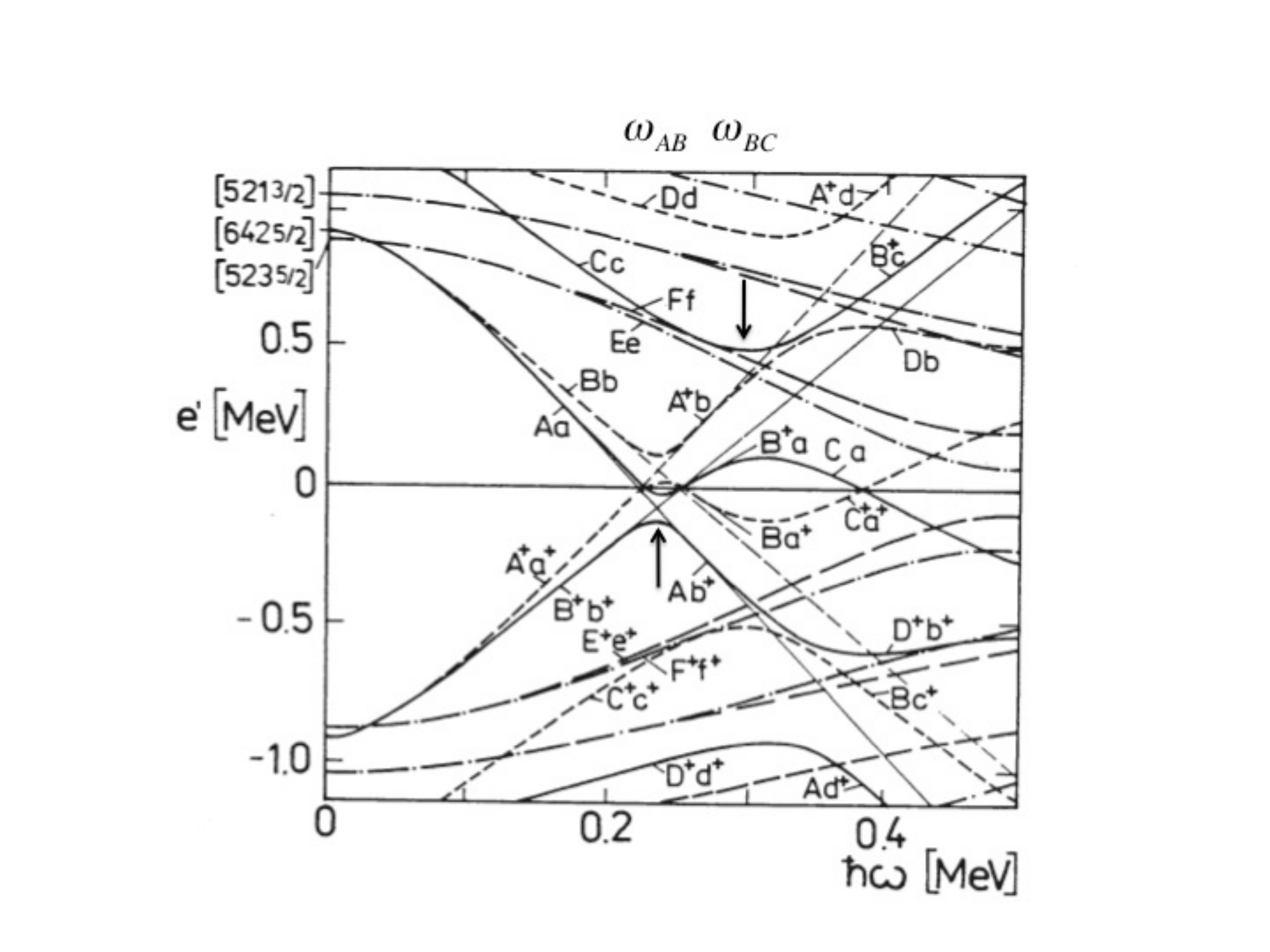}
\vspace*{-1.0cm}
\caption{Quasineutron Routhians for $N\approx 94$. The line type for $(\pi,\alpha)$ is: 
$(+,1/2)$ full, $(+,-1/2)$ short dash, $(-,1/2)$ dash dot, $(-,-1/2)$ long dash. After Ref.\cite{BFqp}.
  }
\label{f:spa}
\end{minipage}
\hspace{0.02\linewidth}
\begin{minipage}[b]{0.38\linewidth}
\hspace*{-1.0cm}
\includegraphics[scale=0.25]{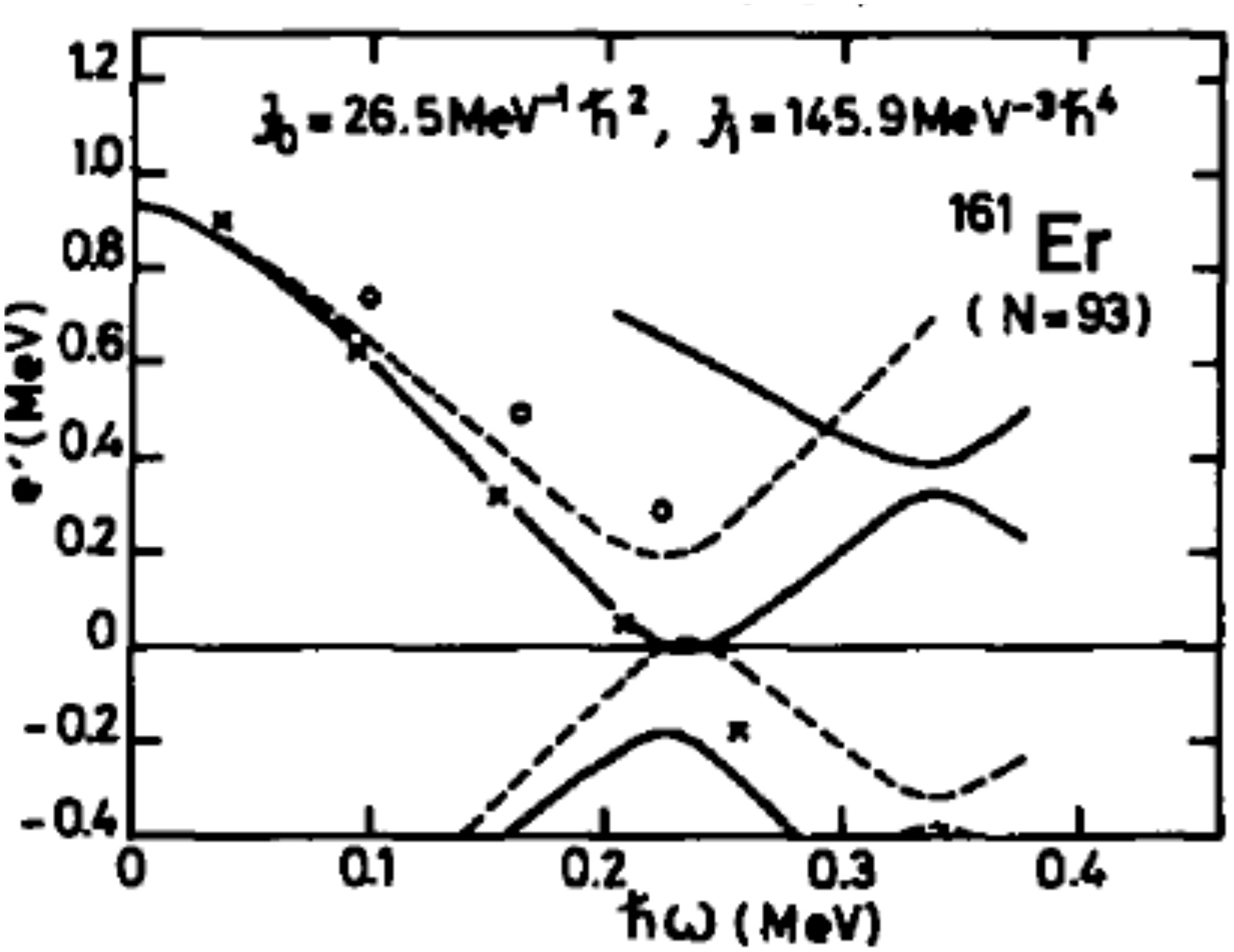}
\vspace*{-0.5cm}
\centering
\caption{Experimental one \newline  quasineutron  Routhians (crosses and circles) compared to CSM calculations. From Ref.\cite{BFqp}.}
\label{f:1qp}
\end{minipage}
\end{figure}

 BF obtained the qp  energies by numerically solving the eigenvalue problem (\ref{e:qpEvEq}) for a fixed monopole pair field
$\Delta_{\alpha i -\alpha k}=\Delta \delta_{i k}$ and using the modified oscillator potential for $h_{sp}$.   Fig. \ref{f:spa} shows an example of qp 
energies in the rotating frame as functions of the rotational frequency $\omega$, which are referred to as qp  Routhians $e'(\omega)$. 
The slope of the qp Routhian indicates the amount of angular momentum $i$ aligned with the rotational axis (alignment $i=-de'/d\omega$, c.f. Eq. (\ref{e:can})).
The  line types indicate their parity $\pi$ and the signature $\alpha$  as $(\pi,\alpha)$, which has become a common notation. 
Each experimental rotational band is assigned to a specific qp configuration. The total signature $\alpha$ is the  sum
of the signatures of  the excited qps. It restricts  the spins of the rotational states to  $\alpha=I$ + even number.   
Attaching letters A, B, ... to the qp  Routhians has become customary for specifying  a multi qp  configuration in a compact way.  
(Conjugate qp states are labeled as A$^+$,  B$^+$,... .  )

Consider the region $\hbar \omega<0.2$ MeV in Fig. \ref{f:spa}. The qp vacuum $\vert 0\rangle$ corresponds to all qp trajectories
$e'(\omega)>0$ unoccupied. It represents the ground state  (g-)  band in the even-even nucleus (e. g. with $N$ =94). The one-q p configurations 
are denoted by A, B, ..., indicating which of the qp trajectories is occupied. For example, the one-quasineutron configurations
$\vert A\rangle=\beta_A^+\vert 0\rangle$
and $\vert B\rangle=\beta_B^+\vert 0\rangle$ represent the $(+,1/2)$  and $(+,-1/2)$ bands in the odd $N=95$ neighboring nuclides. 
Two-qp configurations are denoted by the letters of two occupied trajectories. For example, the two-quasineutron configurations    
 $\vert AB\rangle=\beta_A^+\beta_B^+\vert 0\rangle$
and $\vert AE\rangle=\beta_A^+\beta_E^+\vert 0\rangle$ represent the $(+,0)$  and $(-,1)$ bands in the even $N=94$ nucleus. 
Fig. \ref{f:1qp} compares  the experimental 
 one-quasineutron Routhians (e. g. $e'_A(\omega)=E'_{A}(\omega)-E'_g(\omega)$) with the CSM Routhians. 
 Other one- and two-qp Routhians are equally well reproduced  for many nuclides in different mass regions.

For $\hbar \omega>0.2$ MeV the qp spectrum in Fig. \ref{f:spa} is characterized by avoided crossings between trajectories originating
from regions $e'<0$ and $e'>0$. The first is the "AB" crossing between the trajectories A and B$^+$ at $\hbar \omega_{AB}=0.23$ MeV (marked by an arrow).
These crossings between the qp Routhians allowed BF interpreting the observed irregularities of rotational energies (backbends) in a simple systematic way
as crossing between rotational bands that correspond to different qp configurations.  
 The interpretation of band crossings becomes particularly  transparent by the concept of diabatic qp Routhians, which are constructed by
ignoring the repulsion between crossing qp trajectories, as e. g. the trajectory A in Fig. \ref{f:spa} (shown as a thin line)
continues smoothly through the AB crossing to negative energy. The diabatic
Routhians are labeled by capital letters. 
(Lower case letters denote the adiabatic trajectories that continuously trace  $e'(\omega)$ through the crossings.)

\begin{figure}[t]
\begin{minipage}[b]{0.48\linewidth}
\centering
\hspace*{-0.5cm}
\includegraphics[scale=0.25]{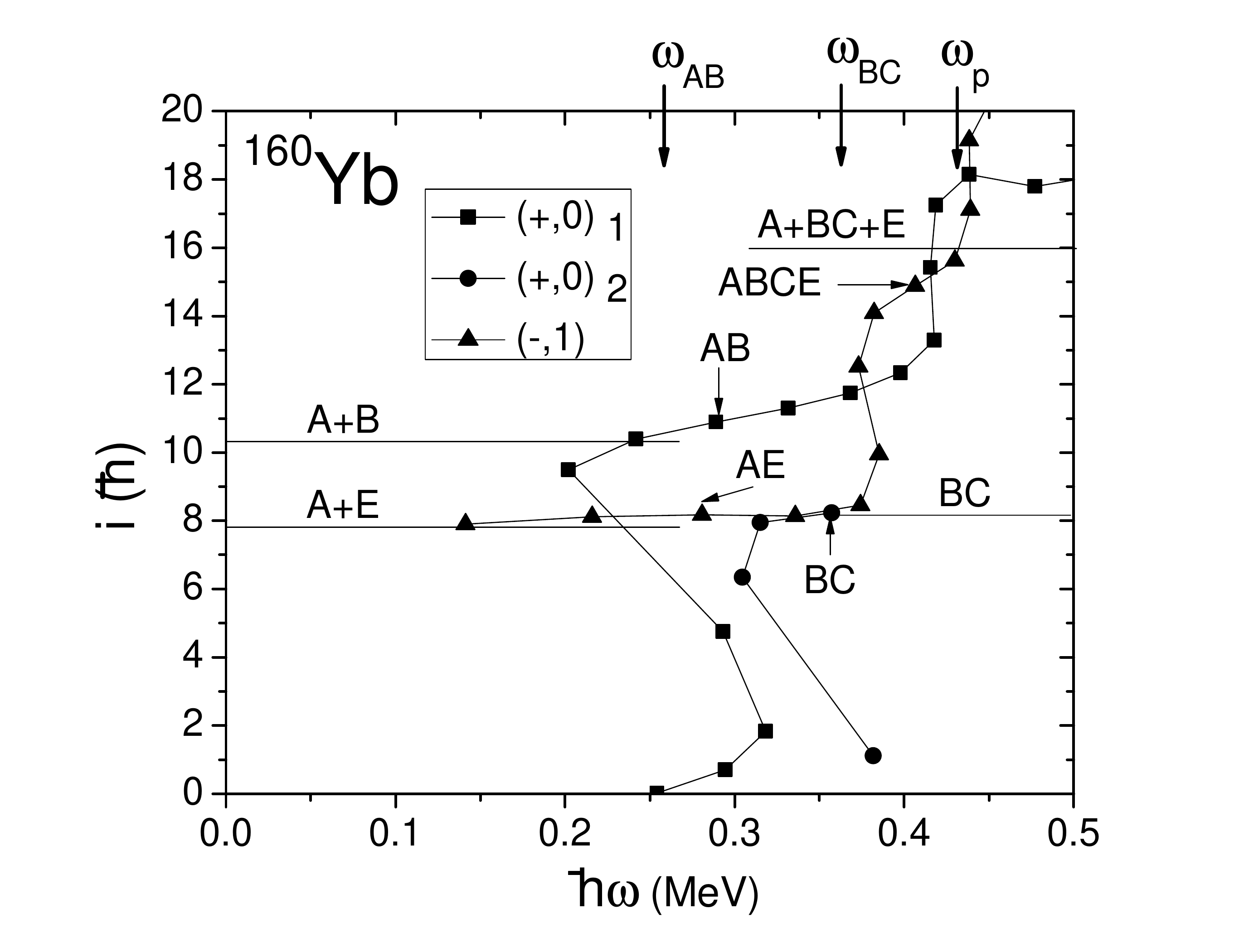}
\centering
\caption{Alignments of multi quasineutron bands in $^{160}$Yb. The horizontal lines
show the values obtained by summing the contributions from the quasiparticle constituents, 
which are indicated by the letters. The reference parameters are ${\cal J}_0=17~\hbar^2$ MeV$^{-1}$ 
and  ${\cal J}_1=70~ \hbar^4 $MeV$^{-3}$. After Ref. \cite{Fint}. }
\label{f:yb160}
\end{minipage}
\hspace{0.02\linewidth}
\begin{minipage}[b]{0.48\linewidth}
\hspace*{-0.7cm}
\includegraphics[scale=0.25]{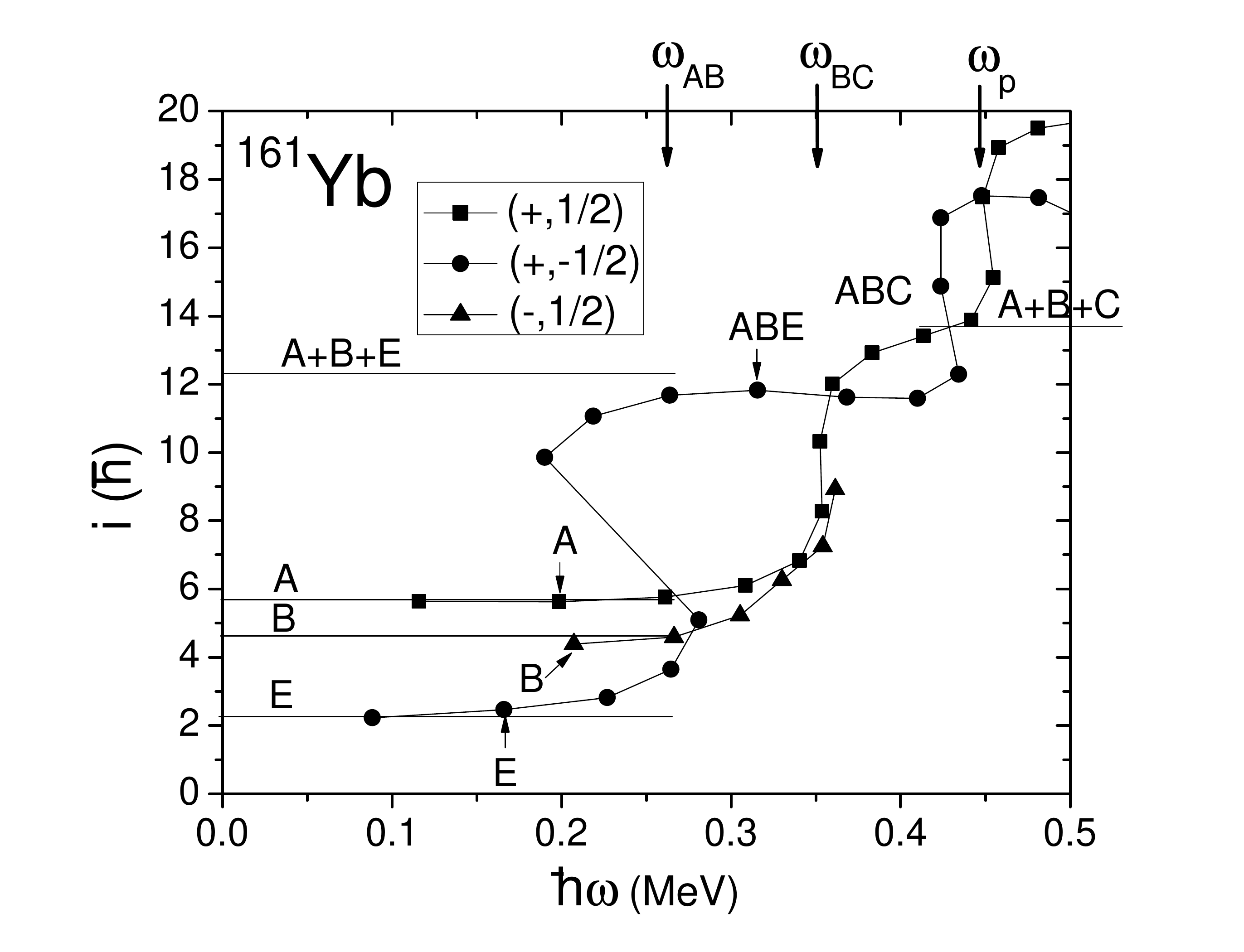}
\caption{Alignments of multi quasi neutron bands in $^{161}$Yb. The horizontal lines
show the values obtained by summing the contributions from the quasi particle constituents, 
which are indicated by the letters. The reference parameters are ${\cal J}_0=19.5~\hbar^2 $ MeV$^{-1}$ 
and  ${\cal J}_1=64~\hbar^4  $MeV$^{-3}$. After Ref. \cite{Fint}.}
\label{f:yb161}
\end{minipage}
\end{figure}

Consider the AB crossing. 
The vacuum $\vert 0\rangle$ has A,B free and  A$^+$, B$^+$ occupied.  It represents the g- band.  The two-quasineutron configuration
$\vert AB\rangle$ has A,B occupied and  A$^+$, B$^+$ free. It represents the s- (Stockholm) band. Before the crossing  the g-band is below the s- band,
and after the crossing the s- band is below the g- band. The experiment exposes the lower (yrast) of the two crossing bands, which changes
from  $\vert 0\rangle$ to $\vert AB\rangle$. The structure change  is observed as the "backbending" effect. 
Fig. \ref{f:yb160}  shows the alignment for the yrast line in $^{160}$Yb. 
The backbend shows around $\omega_{AB}$ as the sudden increase of $i$ by approximately
11$\hbar$, which reflects the 
change from the 0 to the AB configuration. The calculated slopes of A and B correspond to 
 $i_A=6\hbar$ and $i_B=5\hbar$, which add to $i_{AB}=11\hbar$. The large alignments reflects the nearly pure i$_{13/2}$ character of the qps.
 The g- and s- bands exist 
as two bands   in the same nucleus. Sometimes the higher of the two bands is observed too.
 
However, the AB crossing is "blocked"  for the one-quasineutron configuration $\vert A\rangle$ because  both trajectories A and B$^+$ 
are occupied and there is no structural change at the crossing. As seen in Fig. \ref{f:1qp}, the relative Routhian $e'_A=E'_A-E'_g$
goes smoothly through $\omega_{AB}$ becoming negative for $\omega>\omega_{AB}$, and 
the alignment of configuration $\vert A\rangle$ in Fig. \ref{f:yb161} does not change at $\omega_{AB}$. 
The next crossing at $\hbar\omega_{BC}=0.29$ MeV
in Fig. \ref{f:spa} (marked by an arrow)
is not blocked in  configuration $\vert A\rangle$ and indeed seen at $\omega_{BC}$ in Fig. \ref{f:yb161}.
Stephens and Lee and Ring explain the backbending  phenomenon
in more detail in their chapters. 

The CSM premise is that the lowest rotational bands can be constructed by placing qps into one and the same set of levels  which simply add their
contributions. The experimental rotational spectra bear out approximate additivity, which turned out to be a powerful tool for classifying the observed bands.
One consequence of additivity is the appearance of characteristic  band crossing frequencies.  
If some "spectator" qp is added to the two crossing configurations their crossing frequency should not be changed.  For example, the AB crossing 
will appear at about the same $\omega_{AB}$ in all configurations that do not contain
A or B. These are $\vert 0\rangle$ and $\vert E\rangle$ in Figs. \ref{f:yb160} and \ref{f:yb161}.  The AB crossing  
is blocked in configurations that contain A or B, as $\vert A\rangle$ and $\vert B\rangle$ in the figures. The BC crossing at $\hbar\omega_{BC}=0.29$ MeV appears
in Figs. \ref{f:yb160} and \ref{f:yb161} in the configurations  $\vert A\rangle$ 
(which changes to $\vert ABC\rangle)$,  $\vert AE\rangle$ (which changes to $\vert ABCE\rangle)$, and $\vert 0\rangle$  (which changes to $\vert BC\rangle$).
The BC crossing is blocked in $\vert AB\rangle$ and $\vert ABE\rangle$. The same pattern of crossings is expected when a quasiproton spectator is added, which is
indeed observed. 
All configurations in in Figs. \ref{f:yb160} and \ref{f:yb161} show an upbend at $\omega_p$, which is caused by  a 
crossing involving  the h$_{11/2}$ protons. This crossing is expected to be blocked if the additional quasiproton has  h$_{11/2}$ character,
which is observed and was used to identify the nature of the  $\omega_p$ crossing (see Stephens and Lee).  

The CSM became so popular with experimentalists because, as for other versions of the shell model, one can take the qp Routhians and alignments
from experiment for predicting their values in multi qp configurations.   Figs. \ref{f:yb160} and \ref{f:yb161} demonstrate this for the aligned angular momentum $i$.
The lines display the values obtained by adding the contributions of the indicated constituents. 
The remarkable agreement with the observed values is generally found and has become an indispensable tool for identifying  the  qp configurations
of rotational  bands. Frauendorf {\it et al.} \cite{Fint}
went one step further by assuming binary interaction matrix elements between the qps, which were obtained from the deviations of the observed Routhians from
the sum of their constituents. They found matrix elements ranging from -300 to 50 keV, which  increase linearly with $\omega$. 
The change of the pair field alone (cf. section \ref{s:trans}) generates  matrix elements of the order of $ -100$ keV, which could not
explain the size and the state dependence of the experimental ones.    
 \begin{figure}[t]
\begin{minipage}[b]{0.58\linewidth}
\centering
\hspace*{-0.8cm}
\includegraphics[scale=0.29]{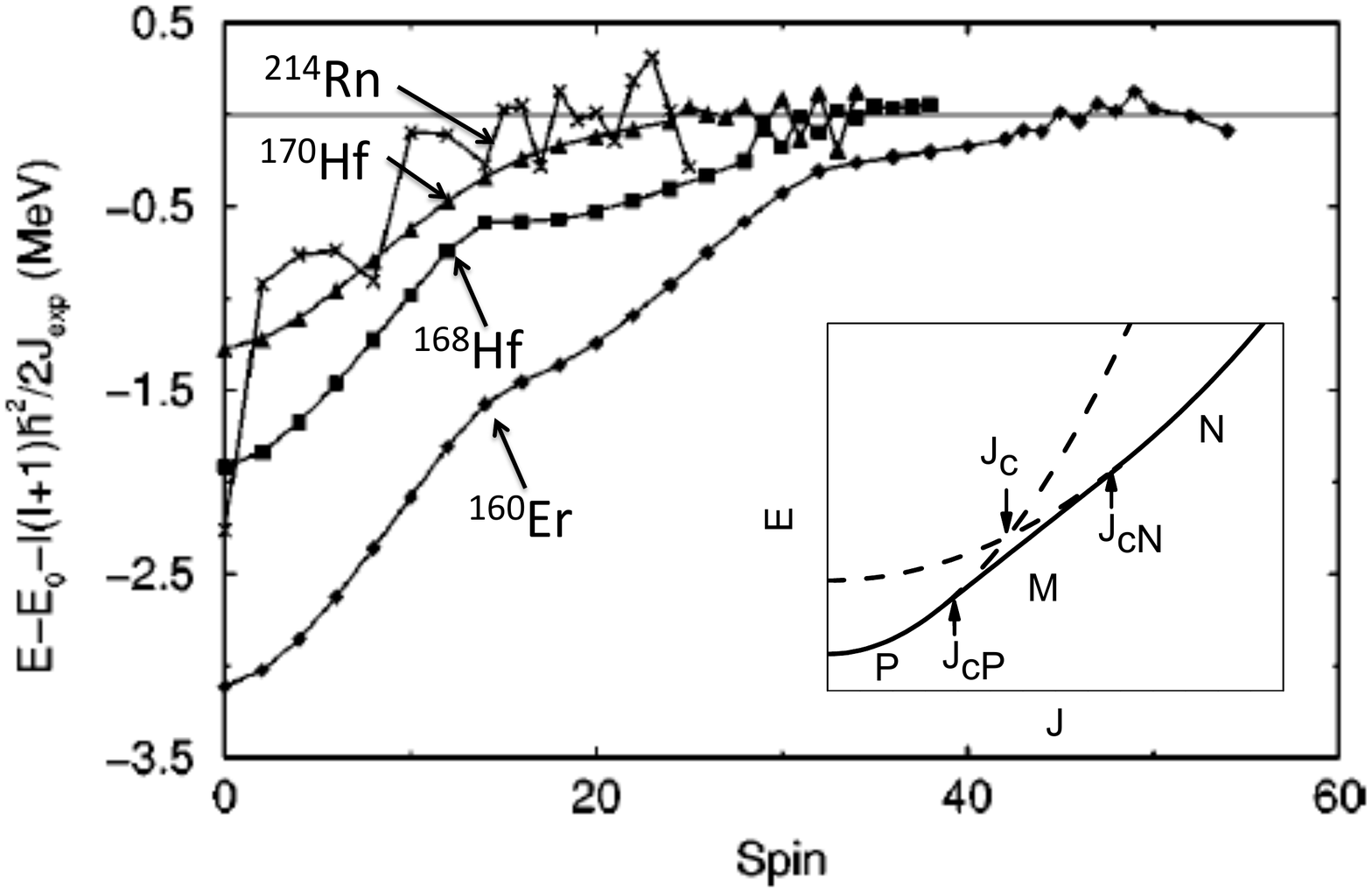}
\vspace*{-1.2cm}
\caption{Difference between the yrast energy and an unpaired rotor. The values 
${\cal J}_{rig}={\cal J}_N= 61,~58,~56,~56~\hbar^2$MeV$^{-1}$ for $^{160}$Er, $^{168}$Hf, $^{170}$Hf, and $^{214}$Rn, respectively.
 From Ref. \cite{MoInop}. Inset: First order phase transition from the
paired (P) phase via the mixed phase (M) to the normal phase (N).}
\label{f:MoInop}
\end{minipage}
\hspace{0.02\linewidth}
\begin{minipage}[b]{0.38\linewidth}
\hspace*{-1.0cm}
\includegraphics[scale=0.38]{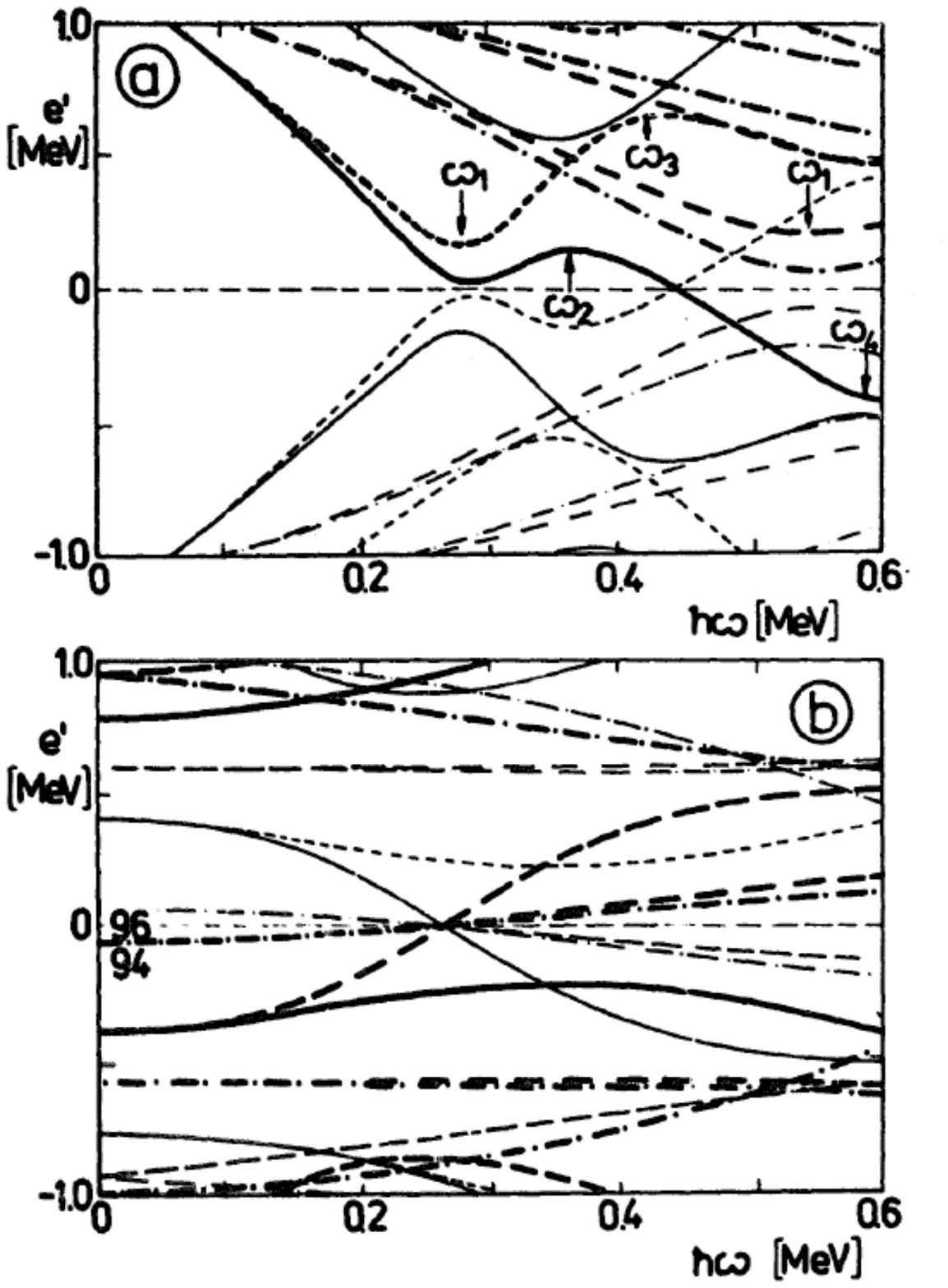}
\vspace*{-1.5cm}
\centering
\caption{a: Quasineutron routhians for $\Delta=1.2$ MeV. b: Particle (thick) and hole (thin) Routhians for  $\Delta=0$ (b).  
From Ref. \cite{Ftrans}.}
\label{f:qpsp}
\end{minipage}
\end{figure}

The CSM rather well predicts  the relative energies and alignments of the various qp configurations. However, it 
it fails at the avoided crossings.  The problems are discussed in Ring's chapter. BF went around them  by resorting to diabatic qp trajectories. 
In their approach, the 
non-interacting g- and s- bands cross each other at the critical angular momentum $J_{AB}$, where the alignment jumps from 0 to $i_s$. 
Because at the crossing $J_g=J_s=J_{AB}$, $\omega_s$ must be smaller than $\omega_g$ in order  to compensate for the gain $i_s$, which is 
the backbend of the yrast states (+,0).    Fig. \ref{f:yb160} shows a more gradual transition from the g- to s- band instead  of the sharp jump.
In other nuclides the transition is even more gradual like the up bends at $\omega_{BC}$ and $\omega_p$ in Figs. \ref{f:yb160} and \ref{f:yb161}. 
The smoothing of the jump is caused by interaction 
between states of the same $I$ in the two bands.   Bengtsson and Frauendorf \cite{BFint} quantitatively related the strength of this interaction to
the repulsion between the trajectories A and B$^+$.

 
\section{Transition to the normal state}\label{s:trans}  

Equating the energies of the paired with the unpaired phases by means of  
 Eq. (\ref{e:omc}), one finds  $\hbar\omega_c=0.28$MeV, $J_{cP}=8.5\hbar$, and
   $J_{cN}=22.6\hbar$ for $E_N-E_P= 2$MeV, ${\cal J}_P= 30 \hbar^2$MeV$^{-1}$, ${\cal J}_N= 80 \hbar^2$MeV$^{-1}$, which are typical for rare earth nuclei.
Deleplanque {\em et al.}\cite{MoInop} inspected the yrast sequences of all even-even nuclei with $A>40$. They observed an approach to  $E(I)\approx I(I+1)/2{\cal J}_N$ for sufficiently high spin, which they interpreted as the unpaired regime. The derived experimental values of ${\cal J}_N$ turned out to deviate substantially from the classical
rigid body value. Zero pairing cranking calculations reproduced the experiment. They explained the deviations from the classical rigid body value as a manifestation
of the shell structure.  Fig. \ref{f:MoInop} shows examples of the difference between $E(I)$ and the energy of the unpaired rotor, i. e. the pair correlation energy. The 
considerable variation of the ground state correlation energy among different nuclei was found to be a general phenomenon.  As expected, the transition appears as a
gradual crossover, which is superimposed by irregularities that are caused by the individual reaction of the nucleonic orbitals to the rotation. 
  Nevertheless, the curves show some reminiscence 
with a first order phase transition. At $I_{cP}$  the system enters the mixed state, where   $E(I)$ increases linearly as $\hbar\omega I$ (see inset of Fig. \ref{f:MoInop}).
In a plot of the correlation energy as Fig. \ref{f:MoInop} this  shows up as a change of the sign of the curvature.   The curves show a global tendency
to concave behavior above $I=10$, which is consistent with   the above estimate $I_{cP}\sim 8$.  Above $I=20$, the correlation energies become small, which is
consistent with the estimate $I_{cN}\sim 20$. Fig. \ref{f:yb166}e shows the correlation energy in the rotating frame. The arrow marks the crossing of
 the $(+,0)$ g- band with the (-,1) band, which is unpaired at this frequency.   The frequency of $\hbar \omega_c=0.32$ MeV
  is consistent with the above estimate of the critical frequency $\hbar \omega_c\sim 0.28$ MeV.   

Mottelson and Valatin \cite{MottValat} first estimated the critical frequency by evaluating the selfconsistency
relation (\ref{e:ScDel}). Using a  rough approximation they found $\hbar\omega_c=0.14$ MeV.  From their numerical solutions of the CHFB problem, Banerjee, Mang, and Ring \cite{Banerjee} found that  $\Delta_{n}=0$ for $\omega>\hbar\omega_c=0.22,~0.3$ MeV, for $^{168}$Yb, $^{162}$Er, respectively. Their pioneering work was followed by many other CHFB calculations. Fig. \ref{f:yb166}a shows a typical example. Consider the (+,0) yrast levels. At low frequency, $\Delta_n$ decreases slowly, which is the CAP
within the g-band (c.f  section \ref{s:MoI}). The sudden drop indicates the AB crossing with the s-band ($\omega_1$ in Fig. \ref{f:qpsp}). 
The two quasineutrons block the levels A and B from the pair correlations,
which suddenly reduces  $\Delta_n$ (cf. Stephens and Lee in this book). 
The remaining weak pair field is quickly destroyed by the combination of CAP and the encounter of the crossing 
at $\omega_4$ in Fig \ref{f:qpsp}.   Consider the (-,1) two-quasineutron band AE. 
It starts with a reduced $\Delta_n$ because 
 A and E are blocked. When the band encounters the BC crossing, the blocking of B and C completely destroys the pair field. For the one-quasineutron bands A and E, 
 $\Delta_n$ drops from a reduced value to zero when the respective BC or AB crossings ($\omega_1$ and $\omega_2$ in Fig. \ref{f:qpsp}) are encountered. 
In summary,  rotation destroys the pair field by suddenly breaking individual pairs of nucleons on high-j orbitals, which is observed as band crossings (backbends), combined with
 the CAP, which originates from the reaction of many low-j orbitals to the inertial forces. This is a general result of all CHFB calculations
 (e. g.  Ref. \cite{Banerjee,Goodman,Fleckner,Cwiok,Bengtsson}).    
  Fig. {\ref{f:yb166} shows that the calculated CHFB Routhians (c) already rather well reproduce the experimental ones (e).

 \begin{figure}[t]
\centering
\hspace*{-1.2cm}
\includegraphics[scale=0.48]{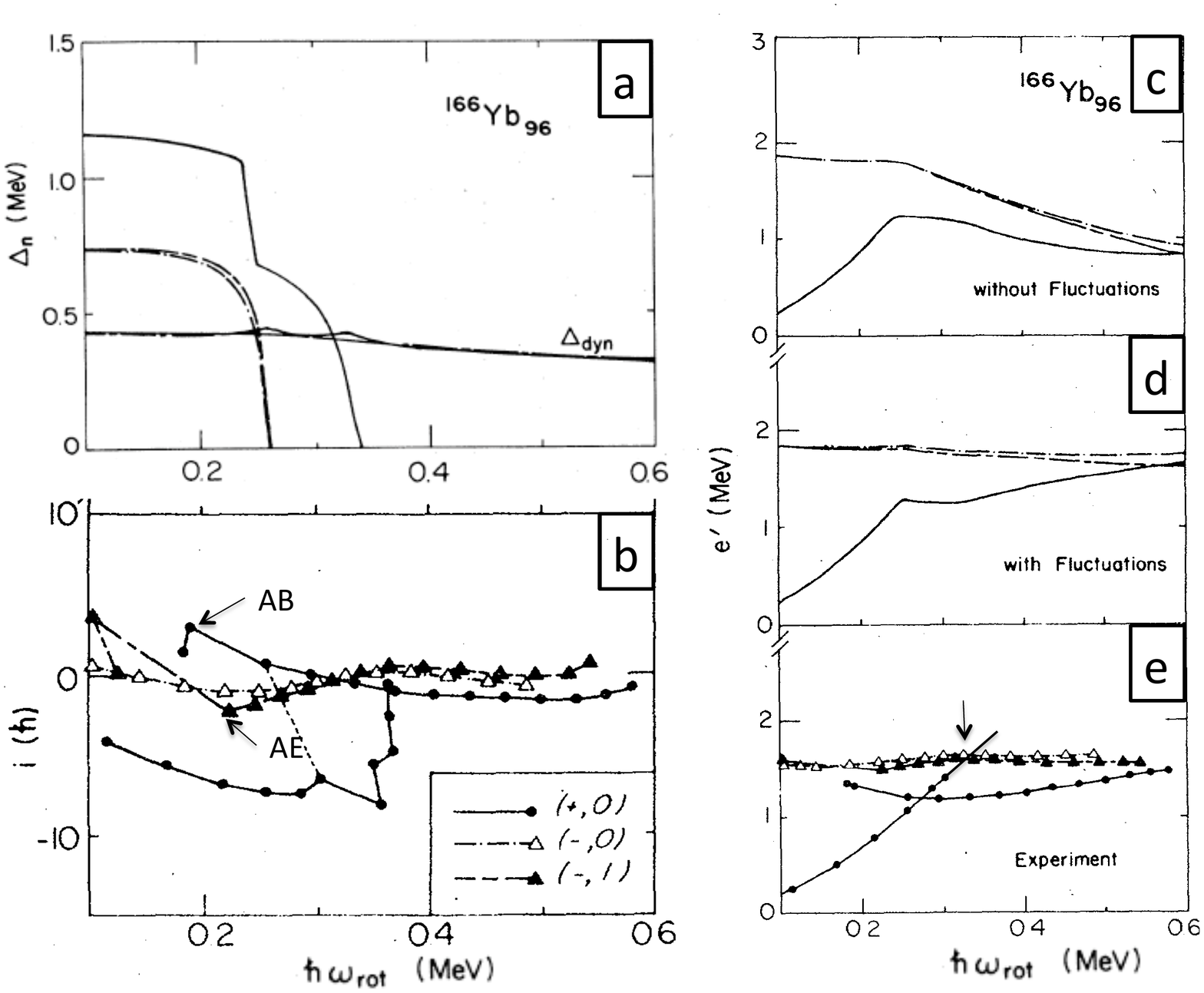}
\vspace*{-1.0cm}
\caption{Three lowest bands in $^{166}$Yb. The line convention for $(\pi,\alpha)$ in b is the same in all panels.
a: Static pair gap $\Delta_n$ calculated from the self consistency condition (\ref{e:ScDel}) using a monopole pairing
interaction (no label). Dynamic pair gap $\Delta_{dyn}=\sqrt{-E_{corr}G}$.  b: Experimental angular momentum
relative to an unpaired rotor $i(\omega)=J(\omega)-{\cal J}_N\omega$. c, d, e: Calculated and experimental Routhians 
relative to an unpaired rotor, $e'(\omega)=E'(\omega)+\omega^2 {\cal J}_N/2$.
The reference parameter  ${\cal J}_0=66, ~62~\hbar^2 $ MeV$^{-1}$
for the experimental and calculated values, respectively.  From Ref. \cite{ShimizuRev}.}
\label{f:yb166}
\end{figure}

The disappearance of the pair field leads to a distinct restructuring of the excitation spectrum with clearly observable consequences that 
have been pointed out by Frauendorf \cite{Ftrans} and Garrett {\em et al.} \cite{GarrettRev}.
 Fig. \ref{f:qpsp} compares the quasineutron Routhians for $\Delta =1.2$ MeV with the ones for $\Delta=0$, which become single particle
 and single hole Routhians.    The following differences are relevant.
 
 \noindent
 1)  The lowest qp Routhians  ( $e'_{\mu\alpha}>0$) have always a negative slope, i. e. a certain amount $i$ of angular momentum is aligned  
 with the rotational axis, which is large for the high-j intruder states and moderate for the normal parity low-j states.  
 The reason is that the lowest qps are composed of comparable particle and hole fractions. This reduces
 the quadrupole moment which binds them to the deformed potential and allows the Coriolis force to partially align the
 qp angular momentum with the rotational axis ("Fermi alignment",  see   \cite{Ftrans}).  As discussed in section \ref{s:CSM},
 the spectra below $\hbar\omega\sim 0.35$ MeV confirm this pattern in a systematic way. 
   For $\Delta=0$ the slopes of the single particle Routhians around the Fermi level change in an erratic way, which is born out by the rotational
   spectra above $\hbar\omega\sim 0.35$ MeV.
 
\noindent
2) The paired regime is characterized by  the systematic appearance of avoided crossings
  between the qp trajectories originating from the negative and positive regions(  $\omega_1, ...,\omega_4$ in Fig. \ref{f:qpsp}a).
 These "pairing induced" crossings are absent in the unpaired regime \cite{Ftrans}, which is confirmed by the high spin
 rotational spectra.  For example,  the backbend in the (+,0) sequence is due to the crossing of the
g-band with the s-band at $\omega_1$. The s-band corresponds
to two-q ps on the lowest levels in Fig . \ref{f:qpsp}a (AB in Fig. \ref{f:spa}). If one follows this configuration
to the case $\Delta=0$ in Fig. \ref{f:qpsp}b it becomes the state with two holes in the lowest N=96 configuration. Since this has N=94,
 the s-band does not exist as an excited configuration in N=96. The coexistence of both g- and s- bands  signifies  the presence of the pair field.
As discussed in section \ref{s:CSM}, the crossings at $\omega_1, ~\omega_2,~\omega_3$ between the $\pi=+$ trajectories are observed systematically.
However, the crossing between the $\pi=-$ trajectories at $\omega_1$ in Fig. \ref{f:qpsp}  (E and F$^+$ in Fig. \ref{f:spa}) is not observed (see Fig. \ref{f:yb166}b),
which indicates the transition from the paired to the unpaired regime around $\hbar\omega=0.35$ MeV.

\noindent
3) The qp spectrum changes in a smooth way with increasing particle number if $\Delta$ is substantially larger than the single particle level distance, 
because the occupation probability gradually changes from 1 to 0 over  a distance of the order $\Delta$. If $\Delta=0$ the erratic single particle spectrum
determines the rotational bands.  The experimental Routhians  of the Yb isotopes  show  the expected transition from a smooth $N$ dependence
 with the characteristic staggering between even and odd $N$ at low $\omega$ to a less regular $N$ dependence with no even-odd staggering at
 $\hbar\omega>0.35$ MeV (see Fig. 13 in Ref. \cite{GarrettRev}).     
 
 It has been generally found that the rotational spectra above some critical frequency $\omega_c$  are accounted for by particle-hole configurations
  built from single particle Routhians like Fig. \ref{f:qpsp}b  (cf. e. g. review articles \cite{AfanasjevRev,SatulaRev}). 
Oliviera {\em et al.} \cite{Oliviera} discussed in detail how the multi qp rotational band spectrum of $^{167,168}$Yb restructures to the  unpaired particle-hole 
spectrum.
For $90\leq N\leq98$, the CHFB calculations predict the disappearance of the neutron pair field between $\hbar\omega=0.3$ and 0.35 MeV, where the restructuring 
of the excitation spectrum is observed.   The transition occurs at lower $\omega$ in the heavier Yb and Hf isotopes \cite{Venkova} and likely also in the 
neutron system of the actinides.   

\section{Fluctuations of the pair field}

The small number of nucleons involved in the correlations implies that the pair field (order parameter) strongly fluctuates around its mean value:
The fixed particle number of nuclei corresponds to a complete delocalization of the orientation (gauge angle), and  
the size of the fluctuations is  comparable with the average value even for the most strongly paired nuclei.
 Continuing preceding studies,  Shimizu {\em et al.} investigated the fluctuations
in the framework of the Random Phase Approximation assuming a monopole pairing interaction. RPA elucidates the role of fluctuation in a particularly simple way: 
It starts from the discussed CHFB solution, which represents  the mean value of the pair field and is called "static pairing". The additional RPA correlations
(described  in Shimizu's chapter), which account for harmonic oscillations of the pair field around its static value,   are called "dynamical pairing". 
(The delocalization of the gauge angle is taken into account). To compare its strength with the static pairing the authors introduced 
$\Delta_{dyn}=\sqrt{-E_{corr}G}$, where $E_{corr}$ is the RPA correlation energy.  As seen in Fig. \ref{f:yb166}a, the dynamic pairing is strong and 
decreases only very slowly with $\omega$, remaining strong after the disappearance of the static gap $\Delta$.   Most of the dynamic pairing originates from 
 single particle states far from the Fermi level, which are insensitive to the break down of the static pairing.   The inclusion of dynamical pairing does not
 qualitatively change the structure of the qp spectrum in the paired regime and of the single particle  spectrum in the unpaired regime, where "paired" and "unpaired" 
 refer to the static  $\Delta$. The elementary excitations are "dressed" by the pair vibrations. However, Figs. \ref{f:yb166}c, d, e demonstrate that the
 dynamical pairing is important for a quantitative  description of the  experiment in the unpaired regime.   

The persistence of strong fluctuations makes the  definition of a boundary between a paired and unpaired "phase" a subtle matter. Refs. 
\cite{Ftrans,ShimizuRev,GarrettRev} advocated the existence 
of a substantial static (CHFB) gap  ($\Delta>\Delta_{dyn}$)  to zone the paired phase, because: i) It becomes the  phase transition
 point in the case of large systems. ii) It marks the change from a spectrum of quasiparticle excitations to particle-hole excitations. iii) It is consistent with 
 the shape of the functions $E(I)$ and $E'(\omega)$ for the yrast levels. iv) It is analogous to our familiar concept   of spherical and deformed nuclei, which are
 classified according to their static shape while there are considerable fluctuations around the average deformation.
 
 The RPA breaks down near $\Delta\rightarrow0$. Shimizu {\em et al.} circumvented the problem by interpolating between the 
 safe regions $\Delta>\Delta_{dyn}$ and $\Delta=0$. Numerical diagonalization of the pairing interaction is the rigorous remedy for
 the problems near the transition point, which is discussed in other  chapters of this book.   X. Wu {\it et al.} in Ref. \cite{Wu} and preceding work
 diagonalized the Routhian (\ref{e:routhian}) for a combination of monopole and quadrupole pairing interactions. As expected, the method accounts very well
 for the experimental data throughout the regions of strong, weak, and zero static paring. They introduced the "particle number conserving pairing gap"
  $\tilde \Delta=\sqrt{-\langle V_P\rangle/G}$ as a measure
 for the pair correlation strength, which they found to  gradually decrease with the rotational frequency $\omega$. The same measure was used before by Egido {\it et al.}
 \cite{Egido1} to quantify the pairing strength in the framework of the particle number projected CHFB approach (see Egido's chapter), which also turned 
 out to persist up to large values of $\omega$.     Based on $\tilde\Delta(\omega)$, these authors concluded that the transition to the normal state is smeared out to
 a degree that it cannot be recognized sometimes.   The "gap" $\tilde\Delta$  measures the total pairing strength originating from the static and dynamic pair correlations.
 Its gradual decrease reflects the persistence of dynamical correlations up to high $\omega$ and, thus, is not  inconsistent with the discussed
 definition of the transition based on the static pair gap $\Delta$ (which should not be confused with $\tilde\Delta$).
 
 The residual dynamical pair correlations after the disappearance of the static pair gap   change only weakly with the rotational frequency and 
  the nucleon configuration, which is  the reason why the numerous calculations  that neglect the pair correlations completely account so well 
  for the experiments on high spin states (see e. g. the review articles \cite{AfanasjevRev,SatulaRev}).  
  
\section{Influence of rotation on the nuclear mean field}\label{s:meanfield}

The properties of rotating nuclei are not only determined by the pair correlations, which are the subject of this chapter. Here I can only mention few
things that are relevant to the preceding sections.  More information can be found in the recent review articles by 
 Afanasjev {\em et al.}\cite{AfanasjevRev} and Satula and Wyss\cite{SatulaRev}. 
 The single particle Hamiltonian $h_{sp}$ in the CHFB Eqs. (\ref{e:qpEvEq})  derives  by additional
self-consistency relations from the qp amplitudes and the effective nucleon-nucleon interaction. Banerjee {\em et al.} \cite{Banerjee}  used the simple 
Quadrupole-Quadrupole interaction, which accounts for changes of the quadrupole deformation.  Bengtsson {\em et al.} \cite{Bengtsson} combined 
the BCS treatment of  pairing with the Strutinsky shell-correction method to determine the shape, an approach that had become very successful  for high 
spin studies (in particular after introducing approximate particle number projection \cite{Satula}). 
Goodman\cite{Goodman} and Fleckner {\em et al.}\cite{Fleckner} solved the CHFB equations starting from the G-matrix and the Skyrme interaction,
respectively. These early attempts to base the CHFB equations on a more fundamental level  resulted only in  modest  agreement with experiment.
Later improvements of the Skyrme interaction substantially improved the accuracy of the calculations (cf. Ref.\cite{Duget} and earlier studies cited). 
Egido and Robledo\cite{Egido2} and Afanasjev {\em et al.}\cite{Afanasjev3} applied  the CHFB (with approximate number projection) to the Gogny interaction
and to the Relativistic Mean Field approach, respectively, which turned out to describe the data very well
(cf. Refs.\cite{Afanasjev1,Afanasjev2}). These approaches
use pair interactions that obey local Galilean invariance and satisfy Eq.  (\ref{e:ScDel}), i. e. the Migdal term is included. A new aspect is the occurrence
of substantial time-odd   terms in the     single particle Hamiltonian $h_{sp}$, which become  important when the effective mass of the nucleons deviates from the real one.     

\section{Conclusions}

The microscopic BCS approach to rotating nuclei accounts for the experimental observations, which characterize the nucleus as a small mesoscopic system. 
The moments of inertia indicate that the nucleus is too small to develop proper superfluidity. The rotation induced transition from the paired to the normal state
can be  recognized in the experimental yrast energies and as a change of the excitation spectrum above the yrast line. It
 does not appear as a sharp phase transition but more as a gradual crossover phenomenon, which is superimposed by irregularities. Still a confined interval
 of rotational frequency for the transition can be identified.


\begin{thebibliography}{9}
\bibitem{BohrMottPines} A. Bohr, B. R. Mottelson and D. Pines, Possible analogy between the 
excitation spectra of nuclei and those of the superconducting metallic state,
   \emph{Phys. Rev. } {\bf 110}, 936- 938 (1958).

\bibitem{Belyaev}
   S.~T. Belyaev, The effect of pairing correlations on nuclear properties,
   \emph{Mat. Fys. Medd. Dan. Vid. Selsk.} {\bf 31}(11), 1- 55(1959).

\bibitem{NilssonPrior}
   S.~G. Nilsson and O. Prior, The effect of pairing correlations on the moment of inertia
   and the collective gyromagnetic ration of deformed nuclei,
   \emph{Mat. Fys. Medd. Dan. Vid. Selsk.} {\bf 32}(16), 1- 60(1961).

\bibitem{Migdal}
   A..~B. Migdal, Superfluidity and the moments of inertia of nuclei,
   \emph{Nucl. Phys.} {\bf 13}, 655- 674 (1959).

\bibitem{Hamamoto}
   A. Hamamoto, The effect of the gauge-invariant pairing interaction on the properties of nuclei,
   \emph{Nucl. Phys.} {\bf 232}, 445- 464 (1974).

\bibitem{Marshalek}
   E.~R. Marshalek, Self-Consistent Perturbation of Hartree-Fock-Bogoliubov
Equations and Nuclear Rotational Spectra. II,
   \emph{Phys. Rev.} {\bf 158}, 993- 1010 (1967).
   
 \bibitem{Frauendorf}
   S. Frauendorf, A systematic investigation of the Coriolis antipairing effect in the
   rare earth region including projection onto exact particle number and angular momentum,
   \emph{Nucl. Phys. A} {\bf 263}, 150- 172 (1974)
   
  \bibitem{Banerjee}
   B. Banerjee, H.~J. Mang and P. Ring, Variational calculation of energy spectra
   of rotational nuclei at high spin ,
   \emph{Nucl. Phys. A} {\bf 215}, 366- 382 (1973)

\bibitem{BFqp}
   R. Bengtsson and S. Frauendorf, Quasiparticle spectra near the yrast line,
   \emph{Nucl. Phys. A} {\bf 327}, 139- 171 (1979)

\bibitem{Fint}
   S. Frauendorf {\it et al.}, Experimental estimates of quasiparticle interactions for rotational nuclei ,
   \emph{Nucl. Phys. A} {\bf 431}, 511- 544 (1984)

\bibitem{BFint}
   R. Bengtsson and S. Frauendorf, An interpretation of backbending  in terms of a 
   crossing between the ground state band and an aligned two-quasiparticle band.,
   \emph{Nucl. Phys. A} {\bf 314}, 27- 36 (1979)

   \bibitem{MoInop}
  M. ~A. Deleplanque {\it et al.}, Gross shell structure at high spin in heavy nuclei,
   \emph{Phys. Rev. C} {\bf 69}, 044309-1-21 (2004)
  
\bibitem{MottValat}
   B.~ R. Mottelson and J. G. Valatin , Effect of nuclear  rotation on the pairing correlation,
   \emph{Phys. Rev. Lett.} {\bf 11}, 511- 512 (1960)

  \bibitem{Ftrans}
  S. Frauendorf, Interplay between single particle and collective degrees of
freedom in rapidly rotating nuclei,
   \emph{Nucl. Phys. A} {\bf 409}, 243 -258 (1983)
  

  
       \bibitem{ShimizuRev}
  Y.~R. Shimizu {\it et al.}, Pairing fluctuations in rapidly rotating nuclei,
   \emph{Rev. Mod. Phys.} {\bf 61}, 131-168 (1989)
  
   \bibitem{GarrettRev}
  J.~D. Garrett {\it et al.}, Recent nuclear structure studies in rapidly rotating nuclei,
   \emph{Ann. Rev. Nucl. Part. Sci.} {\bf 36},419-73 (1986)
  
  \bibitem{Goodman}
 A. Goodman, Self-consistent field description of high spin states in rare earth nuclei,
   \emph{Nucl.  Phys. A} {\bf 265 }, 113 -141 (1976)  
    
 \bibitem{Fleckner}
 J. Fleckner {\em et al.}, Self-consistent calculation of rotational states
in the  rare  earth region with  an effective ground state interaction,
   \emph{Nucl.  Phys. A} {\bf 331 }, 288-310 (1979)  
   
 \bibitem{Cwiok}
  S. \'Cwiok {\it et al.}, Analysis of the backbending effect in $^{166}$Yb, $^{168}$Yb, and $^{170}$ Yb within the Hartree-Fock-
Bogolyubov cranking method,
   \emph{Phys. Rev. C} {\bf 21}, 448-452 (1980)
   
    \bibitem{Bengtsson}
 R. Bengtsson {\em et al.}, Deformation changes along the yrast line in $^{160}$Yb,
   \emph{Nucl.  Phys. A} {\bf 504 }, 221-236 (1983)  
      
    \bibitem{Oliviera}
  J.~B.~R. Oliviera {\it et al.}, Rotation-induced transition from superfluid to normal phase
in mesoscopic systems: $^{168}$Yb and adjacent nuclei,
   \emph{Phys. Rev. C} {\bf 47}, R926-R929 (1993)

   \bibitem{Venkova}
  Ts. Venkova {\it et al.}, Suppression of band crossing in the neutron-rich nuclei $^{172,173}$Yb
due to the absence of a static pair  field,
   \emph{Eur. Phys. J. A} {\bf 26}, 19-24 (2005)
 
  \bibitem{Egido1}
  J.~L. Egido {\it et al.}, On the validity of the mean field approach
for the description of pairing collapse in finite nuclei,
   \emph{Phys. Lett. B} {\bf 154},1-5 (1985)

   \bibitem{Wu}
  X. Wu {\it et al.}, Nuclear pairing reduction due to rotation and blocking,
   \emph{Phys. Rev. C} {\bf 83},034323-1-6 (2011)
 
  \bibitem{AfanasjevRev}
A.~V. Afanasjev {\it et al.}, Termination of
rotational bands: disappearance of quantum many-body collectivity,
   \emph{Phys. Rep.} {\bf 322},1-124 (1999)

  \bibitem{SatulaRev}
  W. Satula and R. Wyss, Mean-field description of high-spin states ,
   \emph{Rep. Prog. Phys.} {\bf 68 }, 131-200  (2005)

\bibitem{Duget}
  T. Duget {\em et al.}, Rotational properties of $^{252, 253, 254}$No:
influence of pairing correlations.
   \emph{Nucl. Phys. A} {\bf 679}, 427Ð440 (2001).

\bibitem{Egido2}
 J.~L. Egido and L.~M. Robledo, High-Spin States in Heavy Nuclei with the Density Dependent Gogny Force,
  \emph{Phys. Rev. Lett.} {\bf 70},2876-2879 (1993)

\bibitem{Afanasjev3}
   A.~V. Avanasjev {\em et al.}, Cranked relativistic Hartree-Bogoliubov theory: Superdeformed bands in the A;190 region,
   \emph{Phys. Rev. C} {\bf 60},051303-1-5 (1999).

\bibitem{Afanasjev1}
   A.~V. Afanasjev {\em et al.}, Moments of inertia of nuclei in the rare earth region: A relativistic
versus nonrelativistic investigation
   \emph{Phys. Rev. C} {\bf 62}, 054306-1-7 (2000).

\bibitem{Afanasjev2}
  A.~ V. Afanasjev and H. Abusara, Time-odd mean fields in covariant density functional theory: Rotating systems,
  \emph{Phys. Rev. C} {\bf 81}, 034329-1-20 (2010)
  
\bibitem{Satula}
 W. Satula {\em et al.}, The Lipkin-Nogami formalism for the cranked mean field,
  \emph{Nucl. Phys. A} {\bf 578}, 45-61 (1994)
  
   
 \end{thebibliography}
\end{document}